\documentclass[runningheads]{llncs}
\usepackage[T1]{fontenc}
\usepackage{graphicx}

\usepackage{makecell}
\usepackage{multirow}
\usepackage{amsmath}
\usepackage{mathrsfs}
\usepackage[bookmarks=true]{hyperref}
\usepackage{lipsum}
\usepackage{xcolor}
\usepackage{bbding}

\definecolor{gray}{RGB}{128,128,128}
\SetLipsumParListSurrounders{\begingroup\color{gray}}{\endgroup}
\begin{document}
\title{Boundary-Aware Network for Abdominal Multi-Organ Segmentation}
%
%
\author{
Shishuai Hu
\and
Zehui Liao
\and
Yong Xia\Envelope
} 
\authorrunning{S. Hu et al.}
\institute{National Engineering Laboratory for Integrated Aero-Space-Ground-Ocean Big Data Application Technology, School of Computer Science and Engineering, Northwestern Polytechnical University, Xi’an 710072, China \\
\email{yxia@nwpu.edu.cn}}
\maketitle              
%

\begin{abstract}
Automated abdominal multi-organ segmentation is a crucial yet challenging task in the computer-aided diagnosis of abdominal organ-related diseases.
Although numerous deep learning models have achieved remarkable success in many medical image segmentation tasks, accurate segmentation of abdominal organs remains challenging, due to the varying sizes of abdominal organs and the ambiguous boundaries among them.
In this paper, we propose a boundary-aware network (BA-Net) to segment abdominal organs on CT scans and MRI scans.
This model contains a shared encoder, a boundary decoder, and a segmentation decoder.
The multi-scale deep supervision strategy is adopted on both decoders, which can alleviate the issues caused by variable organ sizes.
The boundary probability maps produced by the boundary decoder at each scale are used as attention to enhance the segmentation feature maps.
We evaluated the BA-Net on the Abdominal Multi-Organ Segmentation (AMOS) Challenge dataset and achieved an average Dice score of 89.29$\%$ for multi-organ segmentation on CT scans and an average Dice score of 71.92$\%$ on MRI scans.
The results demonstrate that BA-Net is superior to nnUNet on both segmentation tasks.

\keywords{Boundary-aware network  \and Medical image segmentation \and Multi-organ segmentation}
\end{abstract}

\section{Introduction}
\begin{figure}[t]
  \centering
  \includegraphics[width=0.9\textwidth]{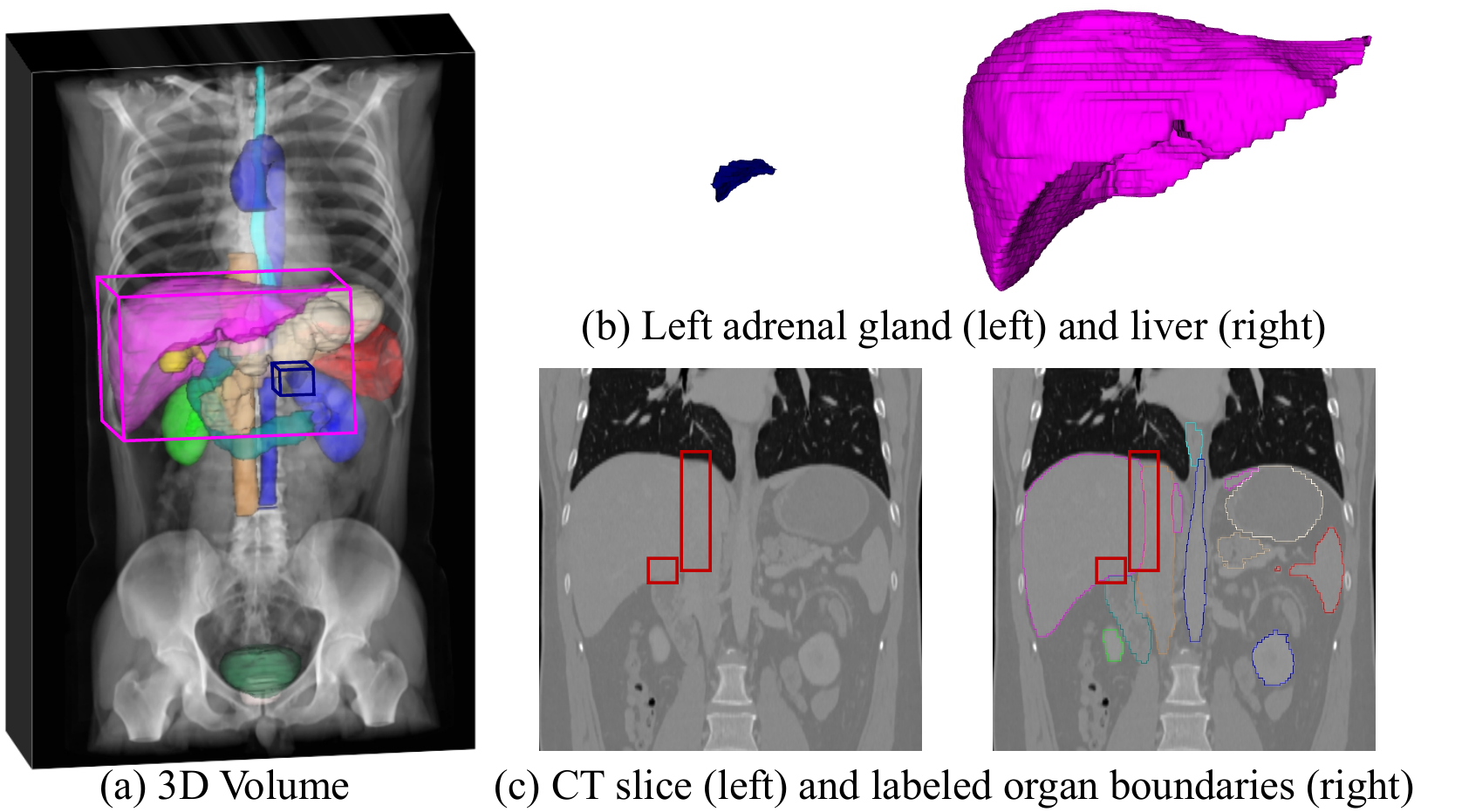}
  \caption{Illustrations of (a) 3D abdominal organ volumes, (b) imbalanced organ volumes, \textit{e.g.}, left adrenal gland and liver, and (c) ambiguous boundaries between different organs and their surroundings.}
  \label{fig:challenges}
\end{figure}

Accurate abdominal anatomy segmentation using computerized tomography (CT) or magnetic resonance imaging (MRI) provides crucial information such as the interrelations among the organs as well as individual positions and shapes in the standardized space, playing an essential role in the computer-aided diagnosis applications such as disease diagnosis and radiotherapy planning~\cite{ji_amos_2022}.
Since manual segmentation of abdominal organs is time-consuming and requires high concentration and expertise, automated segmentation methods are highly demanded to accelerate this process.
However, this task remains challenging due to two reasons: 
(1) abdominal organ regions, $e.g.$, liver and adrenal glands, vary significantly in the volumes, as shown in~\figurename{~\ref{fig:challenges}} (b); and
(2) the contrast between different organs and their anatomical surroundings is particularly low, resulting in the blurry and ambiguous boundary, as shown in~\figurename{~\ref{fig:challenges}} (c).

With the recent success of deep convolutional neural networks (DCNN) in many vision tasks, a number of DCNN-based methods have been proposed for abdominal organ segmentation.
Milletari \textit{et al.}~\cite{milletari2016v} proposed a simple but efficient V-shape network and a Dice loss for volumetric medical image segmentation.
Based on that, Isensee \textit{et al.}~\cite{isensee_nnu-net_2021} integrated the data attributes and network design using empirical knowledge and achieved promising performance on the abdominal organ segmentation task.
Following the same philosophy, Peng \textit{et al.}~\cite{Peng_2022_CVPR} and Huang \textit{et al.}~\cite{huang2022adwunet} searched networks from data directly using NAS and further enhanced the accuracy of abdominal organ segmentation.
Although achieved promising performance, these methods~\cite{milletari2016v,isensee_nnu-net_2021,Peng_2022_CVPR,huang2022adwunet} ignore the boundary information, which is essential for abdominal organ segmentation.
Compared to the largely imbalanced volume proportions of different organs' regions, the boundary ($a.k.a.$ surface in 3D) proportions of different organs are less sensitive to the varying sizes.
Therefore, a lot of boundary-involved segmentation methods have been recently proposed for multi-organ segmentation.
Kervadec \textit{et al.}~\cite{kervadec2019boundary} introduced a boundary loss that uses a distance metric on the space of boundaries for medical image segmentation to enhance the segmentation accuracy near the boundary.
Karimi \textit{et al.}~\cite{karimi2019reducing} proposed a hausdorff distance loss to reduce the hausdorff distance directly to improve the similarity between the predicted mask and ground truth.
Shit \textit{et al.}~\cite{shit2021cldice} incorporated morphological operation into loss function calculation of thin objects, and proposed a clDice loss for accurate boundary segmentation.
Jia \textit{et al.}~\cite{jia2019hd} introduced a 2D auxiliary boundary detection branch that shares the encoder with the 3D segmentation branch and achieved superior performance on MRI prostate segmentation.
Despite the improved performance, the boundary-involved loss function-based methods~\cite{kervadec2019boundary,karimi2019reducing,shit2021cldice} are time-consuming during training since these boundary-involved loss functions are computing-unfriendly, $i.e.$, the distance map of each training image mask should be computed during each iteration.
Whereas the previous auxiliary boundary task-based method~\cite{jia2019hd} does not fully utilize the extracted boundary information since the auxiliary boundary branch will be abandoned at inference time.

In this paper, we propose a boundary-aware network (BA-Net) based on our previous work~\cite{hu_boundary-aware_2020} for abdominal multi-organ segmentation. 
The BA-Net is an encoder-decoder structure~\cite{ronneberger_u-net_2015}.
To force the model to pay more attention to the error-prone boundaries, we introduce an auxiliary boundary decoder to detect target organs' ambiguous boundaries. 
Compared to our previous work in~\cite{hu_boundary-aware_2020}, both the boundary and segmentation decoders are supervised at each scale of the decoder in this paper to further improve the model's robustness to varying sizes of target organs.
Also, we modified the feature fusion mechanism and used the detected boundary probability maps as attention maps on the corresponding segmentation features instead of directly concatenating the boundary feature maps with segmentation feature maps.
We have evaluated the proposed BA-Net model on the AMOS Challenge training dataset using 4-fold cross-validation and achieved a Dice score of 89.29$\%$ on average using CT scans and a Dice score of 71.92$\%$ on average using MRI scans.

\section{Dataset}
The large scale abdominal multi-organ segmentation dataset published by AMOS challenge\footnote{\url{https://amos22.grand-challenge.org/}}~\cite{ji_amos_2022} was used for this study.
The AMOS dataset contains 500 CT and 100 MRI scans with voxel-level annotations of 15 abdominal organs, including the spleen, right kidney, left kidney, gallbladder, esophagus, liver, stomach, aorta, inferior vena cava, pancreas, right adrenal gland, left adrenal gland, duodenum, bladder, and prostate/uterus.
Among all the scans, 200 CT and 40 MRI scans are provided as training cases, 100 CT and 20 MRI scans are used for online validation, and the left 200 CT and 40 MRI scans are withheld for final evaluation. 
Only the voxel-level annotations of training cases are publicly available, while the annotations of validation cases and testing cases cannot be accessed.

\begin{figure}[t]
  \centering
  \includegraphics[width=1\textwidth]{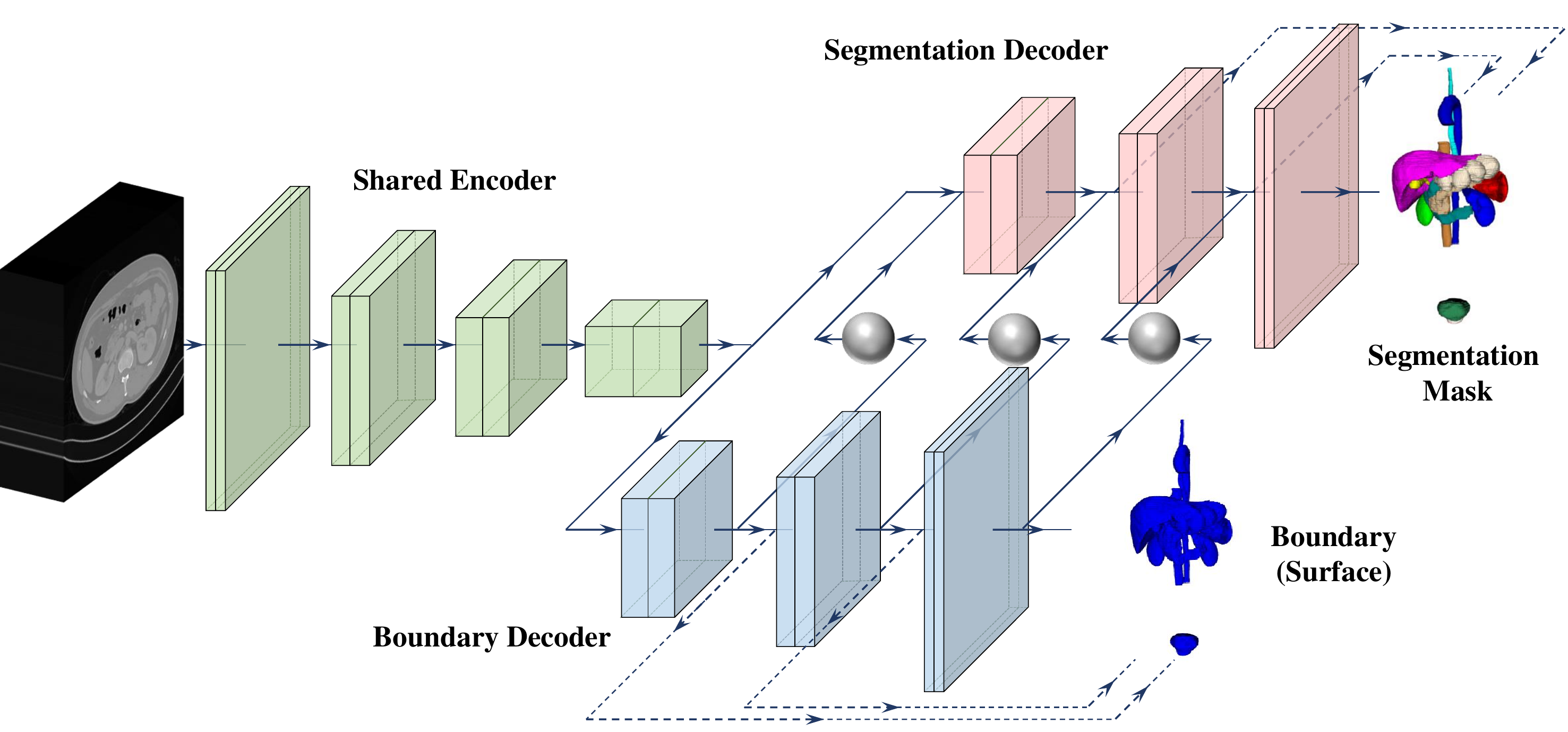}
  \caption{Illustration of BA-Net for abdominal multi-organ segmentation. The green, blue, and pink layers represent convolutional layers of the encoder, boundary decoder, and segmentation decoder respectively. The gray nodes represent the soft-max operations. The dashed lines in both decoders represent deep supervision. The skip connections from the shared encoder to both decoders are omitted for simplicity.}
  \label{fig:overview}
\end{figure}

\section{Method}
The proposed BA-Net contains a shared encoder, a boundary decoder, and a segmentation decoder, as shown in~\figurename{~\ref{fig:overview}}.
The encoder extracts features from the input images. 
Then the boundary decoder performs target organ boundary detection using the extracted features. 
The segmentation decoder takes the extracted features as input and outputs segmentation maps of target organs, wherein the boundary probability maps produced by the boundary decoder are adopted as attention maps to enhance the segmentation features.
\subsection{Shared Encoder}
Since nnUNet~\cite{isensee_nnu-net_2021} performs well on abdominal multi-organ segmentation, we adopt it as the backbone of our BA-Net.
Therefore, the shared encoder design is the same as nnUNet, which is generated from the statistics of the training data.
It contains $N$ encoder blocks ($N$ depends on the specific imaging modalities), each of which is composed of 2 convolutional layers.
Each convolutional layer is followed by instance normalization and the LeakyReLU activation.
The down-sample operations are performed using strided convolutions, followed after each encoder block.
In the encoder, the number of convolutional filters is set to 32 in the first layer, then doubled in each next block, and finally fixed with 320 when it becomes larger than 256.

\subsection{Boundary Decoder}
Symmetrically, the boundary decoder contains $N$ decoder blocks to upsample the feature map extracted by the encoder and gradually refine it to generate the boundary map.
The transposed convolution with a stride of 2 is used to improve the resolution of input feature maps.
The upsampled feature map is concatenated with the low-level feature map extracted from the corresponding encoder block and then fed to the decoder block.
The output feature map $f_i^b$ of each decoder block is processed by a convolutional layer and a soft-max layer to generate the boundary probability map $p_i^b$ at each scale $i$, where $i=1,2,\cdots,N-1$.

\subsection{Segmentation Decoder}
The structure of the segmentation decoder is similar to the boundary decoder, except for an additional step to enhance the segmentation feature map at each scale.
In each block of the segmentation decoder, the upsampled feature map $U(f_{i-1}^s)$ is enhanced with the corresponding boundary probability map $p_i^b$, shown as follows

\begin{equation}
\begin{aligned}
\widehat{U(f_{i-1}^s)} = (1 + p_i^b) * U(f_{i-1}^s)
\end{aligned}
\end{equation}
where $\widehat{U(f_{i-1}^s)}$ is the enhanced feature map, and $U(\cdot)$ represents upsample operation.

The enhanced feature map is concatenated with the feature map extracted from the corresponding encoder block before being further processed by the segmentation decoder block.
Similar to the boundary decoder, the output feature map $f_i^s$ of each segmentation decoder block is also processed by a convolutional layer and a soft-max layer to generate the segmentation probability map $p_i^s$ at each scale $i$.

\subsection{Training and Inference}
The combination of Dice loss and cross-entropy loss is adopted as objective for both the boundary detection and organ segmentation tasks.
The joint loss at each scale can be calculated as

\begin{equation}
\begin{aligned}
\mathcal{L}_i=& 1-\frac{2 \sum_{v=1}^{V} p_{iv} y_{iv}}{\sum_{v=1}^{V}\left(p_{iv}+y_{iv}+\epsilon\right)} \\
&-\sum_{v=1}^{V}\left(y_{iv} \log p_{iv}+\left(1-y_{iv}\right) \log \left(1-p_{iv}\right)\right)
\end{aligned}
\end{equation}
where $p_{iv}$ and $y_{iv}$ denote the prediction and ground truth of the $v$-th voxel in the output of the $i$-th decoder block, $V$ represents the number of voxels, and $\epsilon$ is a smooth factor to avoid dividing by 0.
For the segmentation task, $p_{i}$ is the segmentation probability map $p_i^s$, and $y_i$ is the downsampled segmentation ground truth $y_i^s$. For the boundary detection task, $p_{i}$ is the boundary probability map $p_i^b$, and $y_i$ is the boundary $y_i^b$ of the segmentation ground truth $y_i^s$.

We use deep supervision for both decoders to ensure the boundary can be detected at each scale to enhance the segmentation feature map, and improve the model's robustness to targets' sizes.
Totally, the loss is defined as

\begin{equation}
\begin{aligned}
\mathcal{L} = \sum_{i=1}^{N-1}\omega_i(\mathcal{L}_i^{s} + \mathcal{L}_i^{b})
\end{aligned}
\end{equation}
where $\omega_i$ is a weighting vector that enables higher resolution output to contribute more to the total loss~\cite{isensee_nnu-net_2021}, $\mathcal{L}_i^{s}$ is the segmentation loss, and $\mathcal{L}_i^{b}$ is the boundary detection loss.

During inference, given a test image, the shared encoder extracts features at each scale, and then each boundary decoder block processes these features and produces a boundary probability map.
Based on the extracted features and the boundary probability maps, the segmentation decoder is only required to output the segmentation result of the last decoder block.

\section{Experiments and Results}
\subsection{Implementation Details and Evaluation Metric}
The AMOS challenge contains two tasks: 
(1) segmentation of abdominal organs using only CT scans;
and (2) segmentation of abdominal organs using both CT and MRI scans.

In both tasks, the Hounsfield Unit (HU) values of CT scans were clipped to the range of $[-991, 373]$ according to the data statistics, and then subtracted the mean and divided by the standard deviation.
For task 1, following the nnUNet~\cite{isensee_nnu-net_2021} framework, we trained the 3D low resolution model (\textit{i.e.}, BA-Net-L), which takes the input image with a voxel size of $3.3mm\times1.7mm\times1.7mm$ as input, and the 3D high resolution model (\textit{i.e.}, BA-Net-H), which takes the input image with a voxel size of $2.0mm\times0.7mm\times0.7mm$ as input.
The segmentation results produced by BA-Net-L and BA-Net-H were averaged to produce the final result (\textit{i.e.}, BA-Net-E) for task 1. 
For task 2, the CT scans and the MRI scans were preprocessed by different policies due to their different imaging mechanisms. 
The preprocess workflow and parameters for CT scans were the same as that of task 1. 
The MRI scans were resampled to a uniform voxel size of $2.0mm\times1.2mm\times1.2mm$, and then normalized by subtracting the mean and dividing by the standard deviation.
The preprocessed CT volumes and MRI volumes were fed to a uniform model (\textit{i.e.}, BA-Net-U) randomly so that the model can perform abdominal multi-organ segmentation for both imaging modalities.
The segmentation results produced by BA-Net-U were adopted as the final results for MRI scans.
Whereas the segmentation results produced by BA-Net-L in task 1 and BA-Net-U in task 2 were averaged as the final results (\textit{i.e.}, BA-Net-R) for CT scans in task 2.
Limited to the GPU memory, the size of the input image volume was set to $64\times160\times160$ and the batch size was set to $2$.
The number of down-sampling layers $N$ was set to 5 for both tasks.
The SGD algorithm was used as the optimizer. The initial learning rate $lr_0$ was set to 0.01 and decayed according to $lr = lr_0\times(1-t/T)^{0.9}$, where $t$ is the current epoch and $T=1000$ is the maximum epoch.
The whole framework was implemented in PyTorch and trained using an NVIDIA 2080Ti.

The segmentation performance was measured by the Dice Similarity Coefficient.
All experiments were performed using 4-fold cross-validation with no postprocessing. 
\subsection{Results}

\begin{table}[]
\centering
\caption{Performance (\%) of nnUNet and our BA-Net in abdominal multi-organ segmentation on \textbf{CT} scans of \textbf{task 1}. The best results are highlighted with \textbf{bold}.}
\label{tab:task1}
\setlength\tabcolsep{1pt}
\renewcommand{\arraystretch}{1.2}
\begin{tabular}{c|c|c|c|c|c|c|c|c}
\hline \hline
Methods & Spleen   & \makecell[c]{Right\\Kidney} & \makecell[c]{Left\\Kidney}         & \makecell[c]{Gall\\Bladder}      & Esophagus & Liver   & Stomach         & Arota   \\
\hline
nnUNet-L    & 96.42 & 95.66 & 94.62 & 81.43 & 84.04 & 97.44 & 90.07 & 94.88        \\
\hline
nnUNet-F    & 95.85 & 95.46 & 94.40 & 83.29 & 84.61 & 97.28 & 90.42 & 95.40        \\
\hline
nnUNet-E    & 96.33 & 96.03 & 94.78 & 83.84 & 85.29 & 97.68 & 90.77 & 95.63        \\
\hline
BA-Net-L    & 95.91 & 95.88 & 94.85 & 83.25 & 84.22 & 97.57 & 90.32 & 94.87        \\
\hline
BA-Net-F    & 96.34 & 95.70 & 94.56 & 83.22 & 84.93 & 97.46 & 90.77 & 95.69        \\
\hline
BA-Net-E    & \textbf{96.46} & \textbf{96.25} & \textbf{95.03} & \textbf{84.16} & \textbf{85.46} & \textbf{97.79} & \textbf{91.11} & \textbf{95.75}        \\

\hline \hline
Methods & Postcava & Pancreas     & \makecell[c]{Right\\Adrenal\\Gland}  & \makecell[c]{Left\\Adrenal\\Gland} & Duodenum  & Bladder & \makecell[c]{Prostate\\/Uterus} & \textbf{Average} \\
\hline
nnUNet-L    & 90.41 & 84.41 & 76.39 & 77.26 & 80.73 & 88.99 & 84.80 & 87.84       \\
\hline
nnUNet-F    & 91.04 & 85.97 & 78.77 & 79.87 & 82.06 & 88.69 & 84.14 & 88.48       \\
\hline
nnUNet-E    & 91.38 & 85.94 & 78.77 & 79.83 & 82.30 & 90.09 & 85.77 & 88.96        \\
\hline
BA-Net-L    & 90.54 & 84.89 & 76.41 & 77.16 & 80.78 & 90.27 & 85.35 & 88.15        \\
\hline
BA-Net-F    & 91.42 & \textbf{86.56} & \textbf{79.11} & \textbf{80.06} & 82.54 & 89.67 & 86.65 & 88.98        \\
\hline
BA-Net-E    & \textbf{91.58} & 86.43 & 78.95 & 79.97 & \textbf{82.55} & \textbf{91.13} & \textbf{86.80} & \textbf{89.29}        \\
\hline \hline
\end{tabular}
\end{table}

\begin{table}[]
\centering
\caption{Performance (\%) of nnUNet and our BA-Net in abdominal multi-organ segmentation on \textbf{CT} scans of \textbf{task 2}. The best results are highlighted with \textbf{bold}.}
\label{tab:task2-ct}
\setlength\tabcolsep{1pt}
\renewcommand{\arraystretch}{1.2}
\begin{tabular}{c|c|c|c|c|c|c|c|c}
\hline \hline
Methods & Spleen   & \makecell[c]{Right\\Kidney} & \makecell[c]{Left\\Kidney}         & \makecell[c]{Gall\\Bladder}      & Esophagus & Liver   & Stomach         & Arota   \\
\hline
nnUNet-L    & 96.42 & 95.66 & 94.62 & 81.43 & 84.04 & 97.44 & 90.07 & 94.88        \\
\hline
nnUNet-U    & 95.96 & 95.47 & 94.35 & 83.05 & 84.56 & 97.20 & 90.29 & 95.35        \\
\hline
nnUNet-R    & 96.35 & 96.05 & 94.84 & 83.09 & 85.28 & 97.65 & 90.75 & 95.56        \\
\hline
BA-Net-L    & 95.91 & 95.88 & 94.85 & 83.25 & 84.22 & 97.57 & 90.32 & 94.87        \\
\hline
BA-Net-U    & 96.20 & 95.70 & 94.84 & 83.46 & 84.73 & 97.44 & 90.45 & 95.56        \\
\hline
BA-Net-R    & \textbf{96.43} & \textbf{96.26} & \textbf{95.13} & \textbf{84.27} & \textbf{85.36} & \textbf{97.74} & \textbf{90.98} & \textbf{95.71}        \\

\hline \hline
Methods & Postcava & Pancreas     & \makecell[c]{Right\\Adrenal\\Gland}  & \makecell[c]{Left\\Adrenal\\Gland} & Duodenum  & Bladder & \makecell[c]{Prostate\\/Uterus} & \textbf{Average} \\
\hline
nnUNet-L    & 90.41 & 84.41 & 76.39 & 77.26 & 80.73 & 88.99 & 84.80 & 87.84       \\
\hline
nnUNet-U    & 90.86 & 85.81 & 78.40 & 79.31 & 81.99 & 88.71 & 84.52 & 88.39        \\
\hline
nnUNet-R    & 91.28 & 85.86 & 78.59 & 79.50 & 82.28 & 90.01 & 85.74 & 88.85        \\
\hline
BA-Net-L    & 90.54 & 84.89 & 76.41 & 77.16 & 80.78 & 90.27 & 85.35 & 88.15        \\
\hline
BA-Net-U    & 91.21 & \textbf{86.33} & \textbf{78.59} & 79.45 & 82.06 & 89.70 & 85.53 & 88.75        \\
\hline
BA-Net-R    & \textbf{91.46} & 86.20 & 78.51 & \textbf{79.52} & \textbf{82.29} & \textbf{91.00} & \textbf{86.60} & \textbf{89.16}        \\
\hline \hline
\end{tabular}
\end{table}

\begin{table}[]
\centering
\caption{Performance (\%) of nnUNet, BA-Net, and their variants in abdominal multi-organ segmentation on \textbf{MRI} scans of \textbf{task 2}. The best results are highlighted with \textbf{bold}.}
\label{tab:task2-mr}
\setlength\tabcolsep{0.8pt}
\renewcommand{\arraystretch}{1.2}
\begin{tabular}{c|c|c|c|c|c|c|c|c}
\hline \hline
Methods & Spleen   & \makecell[c]{Right\\Kidney} & \makecell[c]{Left\\Kidney}         & \makecell[c]{Gall\\Bladder}      & Esophagus & Liver   & Stomach         & Arota   \\
\hline
\makecell[c]{nnUNet-F\\(CT$\to$MRI)}    & 55.96 & 39.23 & 62.03 & 26.30 & 29.82 & 36.96 & 49.68 & 65.09 \\
\hline
nnUNet-M    & \textbf{97.14} & 94.93 & 93.87 & 75.27 & 67.07 & 97.96 & 84.95 & 92.10       \\
\hline
nnUNet-U    & 93.59 & 95.36 & 94.80 & \textbf{77.75} & 66.23 & 97.91 & 87.35 & 92.15        \\
\hline
nnUNet-D    & 94.07 & 95.45 & \textbf{95.04} & 77.01 & 66.19 & 97.89 & 87.92 & 92.23        \\
\hline
BA-Net-U    & 93.62 & \textbf{95.59} & 95.02 & 74.58 & \textbf{67.31} & \textbf{98.10} & \textbf{88.12} & \textbf{92.52}        \\
\hline \hline
Methods & Postcava & Pancreas     & \makecell[c]{Right\\Adrenal\\Gland}  & \makecell[c]{Left\\Adrenal\\Gland} & Duodenum  & Bladder & \makecell[c]{Prostate\\/Uterus} & \textbf{Average} \\
\hline
\makecell[c]{nnUNet-F\\(CT$\to$MRI)}    & 51.08 & 40.85 & 20.61 & 34.94 & 34.54 & 0.00 &  0.00 &  36.47  \\
\hline
nnUNet-M    & \textbf{88.73} & 84.82 & 63.43 & 60.51 & 69.13 & 0.00 &  0.00 & 71.33        \\
\hline
nnUNet-U    & 88.08 & 84.65 & \textbf{64.62} & \textbf{63.65} & 71.39 & 0.00 &  0.00 &  71.84        \\
\hline
nnUNet-D    & 88.45 & 84.26 & 64.36 & 62.35 & 72.07 & 0.00 &  0.00 &  71.82        \\
\hline
BA-Net-U    & 88.46 & \textbf{85.17} & 64.54 & 63.04 & \textbf{72.67} & 0.00 &  0.00 &  \textbf{71.92}        \\
\hline \hline
\end{tabular}
\end{table}

We compared the proposed BA-Net with nnUNet~\cite{isensee_nnu-net_2021} on both tasks.
The performance of nnUNet and BA-Net on CT scan segmentation for task 1 is shown in~\tablename{~\ref{tab:task1}}.
It shows that the overall performance of our BA-Net is superior to nnUNet in both the low-resolution model (nnUNet-L v.s. BA-Net-L) and high-resolution model (nnUNet-H v.s. BA-Net-H).
It demonstrates that our BA-Net can effectively improve the segmentation performance on CT scans.
\tablename{~\ref{tab:task1}} also reveals that the averaged performance of our BA-Net (BA-Net-E) achieves the best result, confirming the importance of taking boundary into consideration for model design.

The performance of nnUNet and BA-Net on CT scans segmentation for task 2 is shown in~\tablename{~\ref{tab:task2-ct}}.
It shows that the joint training of CT scans and MRI scans brings no benefit to CT scan segmentation (nnUNet-F v.s. nnUNet-U).
Still, our BA-Net achieves better performance than nnUNet on every organ.

The performance of nnUNet, BA-Net, and their variants on MRI scan segmentation for task 2 is shown in~\tablename{~\ref{tab:task2-mr}}.
We also trained a nnUNet model only on MRI scans, denoted as nnUNet-M.
Based on the uniform nnUNet model for both CT scans and MRI scans, we introduced domain-specific convolution to process the CT volumes and MRI volumes respectively in the first encoder block, denoted as nnUNet-D.
Also, we directly generate the segmentation results for MRI scans using nnUNet-F that trained using CT scans, denoted as nnUNet-F (CT$\to$MRI), to quantize the domain gap between MRI and CT scans.
\tablename{~\ref{tab:task2-mr}} shows that the joint training can improve the segmentation performance on MRI scans (nnUNet-M v.s. nnUNet-U). We believe that plenty of CT data for training improves the model's feature extraction ability, which enhances the segmentation performance on limited MRI scans.
Surprisingly, \tablename{~\ref{tab:task2-mr}} reveals that the domain-specific convolution makes no contribution to the MRI segmentation performance.
We assume that preprocessing CT scans and MRI scans in different ways can effectively reduce the domain gap between CT and MRI (nnUNet-F v.s. nnUNet-U).
It reveals in~\tablename{~\ref{tab:task2-mr}} that our BA-Net still can improve the segmentation performance on MRI scans (nnUNet-U v.s. BA-Net-U).

\section{Conclusion}
In this paper, we propose a BA-Net for abdominal multi-organ segmentation.
It is composed of a shared encoder for feature extraction, a boundary decoder for boundary detection, and a segmentation decoder for target organ segmentation.
At each scale of the boundary decoder and the segmentation decoder, the boundary and segmentation mask at the corresponding scale are used to supervise the predicted probability map.
The predicted boundary map at each decoder block is used as an attention map to enhance the segmentation features.
Experimental results on the AMOS challenge dataset demonstrate that our BA-Net is superior to nnUNet on CT scan segmentation and MRI scan segmentation.

\bibliographystyle{splncs04}
\bibliography{reference}

\end{document}